 \documentclass[iop]{emulateapj}
 \shorttitle{No IMBHs in Globular Clusters}
\shortauthors{Strader \etal~}
\def\etal{{\it et al.}}

\usepackage{subfigure}
\usepackage{amsmath}
\usepackage{graphics}

\begin{document}

\title{No Evidence for Intermediate-Mass Black Holes in Globular Clusters: Strong Constraints from the JVLA}

\author{Jay Strader\altaffilmark{1}, Laura Chomiuk\altaffilmark{2,1}, Thomas J.~Maccarone\altaffilmark{3}, James C.~A.~Miller-Jones\altaffilmark{4}, Anil C.~Seth\altaffilmark{5}, Craig O.~Heinke\altaffilmark{6}, Gregory R.~Sivakoff\altaffilmark{6}}
\email{jstrader@cfa.harvard.edu}

\altaffiltext{1}{Harvard-Smithsonian Center for Astrophysics, Cambridge, MA 02138, USA}
\altaffiltext{2}{National Radio Astronomy Observatory, P.O. Box O, Socorro, NM 87801, USA}
\altaffiltext{3}{School of Physics and Astronomy, University of Southampton, HighÞeld SO17 IBJ, UK}
\altaffiltext{4}{International Centre for Radio Astronomy Research, Curtin University, GPO Box U1987, Perth, WA 6845, Australia}
\altaffiltext{5}{University of Utah, Salt Lake City, UT 84112}
\altaffiltext{6}{Department of Physics, University of Alberta, 4-183 CCIS, Edmonton, AB, T6G 2E1, Canada}

\begin{abstract}

With a goal of searching for accreting intermediate-mass black holes (IMBHs), we report the results of ultra-deep Jansky VLA radio continuum observations of the cores of three Galactic globular clusters: M15, M19, and M22. We reach rms noise levels of 1.5--2.1 $\mu$Jy beam$^{-1}$ at an average frequency of 6 GHz. No sources are observed at the center of any of the clusters. For a conservative set of assumptions about the properties of the accretion, we set $3\sigma$ upper limits on IMBHs from $360-980 M_{\odot}$. These limits are among the most stringent obtained for any globular cluster. They add to a growing body of work that suggests either (a) IMBHs $\ga 1000 M_{\odot}$ are rare in globular clusters, or (b) when present, IMBHs accrete in an extraordinarily inefficient manner.

\end{abstract}


\section{Introduction}

Supermassive black holes reside in the centers of most massive galaxies, and these black holes are expected to have grown from an initial population of ``seed" black holes through gas accretion and mergers (Volonteri \& Perna 2005). This seed population may originate in Population III stars, direct collapse out of dense gas disks, or the cores of young massive star clusters via runaway stellar mergers (e.g., Portegies Zwart \etal~2004). A subset of these star clusters are expected to survive to the present day and form the observed population of old globular clusters (GCs) around galaxies like our Milky Way. This possibility motivates the search for intermediate-mass black holes (IMBHs; $10^{2}$--$10^{4} M_{\odot}$) in Galactic GCs.

The most popular method for IMBH searches in GCs has been dynamical analyses of stars near the cluster center: objects with central IMBHs will have identifiable signatures in their line-of-sight velocity distributions (including high velocity dispersions) near the radius of influence: typically $\sim 0.05$ pc = 1--$2\arcsec$(Gebhardt \etal~2000; Gerssen \etal~2002, 2003). Unfortunately, in many cases the predicted dynamical signature can be mimicked by other effects (such as a population of mass-segregated neutron stars), and the data are fundamentally limited by shot noise due to the small number of stars within the sphere of influence (e.g., Baumgardt \etal~2003; van der Marel \& Anderson 2010).

Analogous to active galactic nuclei, IMBHs can also be detected by their accretion signatures. At low accretion rates, X-rays are produced by the accretion flow and radio continuum emission through synchrotron radiation from the emitted jets. In a GC, the winds of evolved stars provide a continual flux of gas into the intracluster medium; only a small fraction of this material would need to be accreted by a putative IMBH to provide significant accretion luminosity (e.g., Pooley \& Rappaport 2006).

Merloni \etal~(2003) and Falcke \etal~(2004) showed that the X-ray and radio luminosity of both stellar-mass and supermassive accreting BHs in the low/hard state fall on a fundamental plane with BH mass, suggesting that the underlying accretion physics has commonalities at a wide range of masses and accretion rates. This plane offers the possibility to set quantitative constraints on the mass of an accreting BH via X-ray and radio observations. The sense of the relation is  that for more massive BHs and those with lower accretion rates, the ratio of radio to X-ray luminosity increases. 

The commissioning of the Karl~G.~Jansky~VLA (JVLA), with an eventual increase of a factor of 10--80 in the available bandwidth, has provided an extraordinary improvement in sensitivity for observations of faint radio continuum sources (Perley \etal~2011). As a pilot project, we have obtained very deep images of the GCs M19 (NGC~6273) and M22 (NGC~6656) to search for central IMBHs. These clusters were chosen on the basis of their declination, moderately high masses and close distances, and large ratios of core to half-light radii (proposed as one marker for the presence of an IMBH; Trenti \etal~2007). To these GCs we add M15 (NGC~7078), a cluster long discussed as a potential IMBH host (e.g., Gerssen \etal~2002), for which JVLA data were recently obtained for a transient monitoring campaign.

\section{Data}

We observed M19 and M22 with the JVLA over the period 2011 May 19 to 29 as part of the program 10C-109 (P.I.~Chomiuk). Ten hours were spent observing each source in BnA configuration, split among four 2.5 hour blocks. 7.5 hr were spent on source. M15 was observed as part of a monitoring campaign of the X-ray binary M15 X-2 between 2011 May 22 and 2011 August 22, under program 11A-269 (P.I.~Miller-Jones). Data were obtained in the BnA and A configurations, spread across five epochs for a total of 10 hours (5.25 hr on target).

We observed with 2048 MHz of bandwidth and four polarization products, with two separate 1 GHz basebands centered at 5.0 and 6.75 GHz (5.0 and 7.45 GHz in the case of M15). The resolution our M22 and M19 images is $\sim1.4\arcsec \times 0.9\arcsec$, and $\sim0.8\arcsec \times 0.5\arcsec$ for M15. 

Data were reduced using standard routines in CASA and AIPS. Each observing block was edited for bad data and interference and then calibrated. The two calibrated basebands were split off from one another, and all data for a given object and baseband were concatenated in the \emph{uv} plane and imaged with a Briggs robust weighting of 1. There is a bright extended source near M15's core (the planetary nebula K648), so we self-calibrated these data to minimize sidelobes. To obtain the deepest combined image possible for each GC, we smoothed the image from the higher-frequency baseband to the resolution of the 5.0 GHz image and averaged these two images together, providing an rms sensitivity of 1.5 $\mu$Jy beam$^{-1}$ for M19 and M22, and 2.1 $\mu$Jy beam$^{-1}$ for M15.

\section{Results}

\subsection{Cluster Centers}

Our principal result is that there is no significant radio continuum source at any position consistent with the photometric center of any of our GCs. For M19 and M22 the closest $5\sigma$ sources to the cluster centers are at respective angular separations of 28\arcsec\ and 16\arcsec. M15 has three sources close to the center, and all are known: the X-ray binaries AC 211 and M15 X-2 (Johnston \etal~1991; Miller-Jones \etal~2011) and the pulsar PSR B2127+11A (Wolszczan \etal~1989). The closest unidentified source to the photometric center of M15 is at $\sim~11\arcsec$.

For M15 and M22 the centers are from ellipse fitting of resolved star counts in HST/ACS data, with uncertainties of 0.2\arcsec\ and 0.8\arcsec\ (Goldsbury \etal~2010). The M19 center (uncertainty $\sim 1.1\arcsec$) is from Picard \& Johnston (1995), from ground-based imaging.

The uncertainty in the photometric center is only one component of the overall uncertainty in the possible location of a central IMBH. Encounters with passing stars will perturb the IMBH from the center, and in the mass regime under consideration the perturbations can be significant. For an assumed Plummer model, Chatterjee \etal~(2002; see also van der Marel \& Anderson 2010) predict that the variance of mean-squared one-dimensional deviations will be: $<x^2~>=(2/9) (m_{*}/M_{\rm BH}) r_c^2$ where $m_{*}=<m^2>/<m>$ over the stars in the core, $M_{\rm BH}$ is the IMBH mass, and $r_c$ is the Plummer core radius. Because mass segregation can deplete the core of low-mass stars, we adopt a conservative value of $m_{*}=1 M_{\odot}$. For $M_{\rm BH}$, the most conservative assumption is the defined lower mass limit for an IMBH: $M_{\rm BH}=100 M_{\odot}$. For this set of conservative assumptions, the formula above suggests that the typical displacement from the center is $\sim 5$\% of the core radius. For our actual $3\sigma$ IMBH mass limits (see \S 3.3), the displacements are $\la 1$\% of the core radius, and smaller than the uncertainties in the photometric centers.

M22 has by far the largest core radius of the three GCs; McLaughlin \& van der Marel (2005) find a King radius $r_0=85$\arcsec. Using the equation above, we estimate the typical one-dimensional deviation of a $100 M_{\odot}$ IMBH in the core of M22 to be $\sim 4$\arcsec. The cores of M15 and M19 are much smaller than in M22, with predicted deviations of 0.2\arcsec and 1.3\arcsec. In all cases the closest unidentified $5\sigma$ sources to the cluster centers are all at much larger angular separations than these values. Images of the cores of all of the clusters are shown in Figure 1. The previously unknown non-central sources will be discussed in future papers.

We conclude that there are no candidate IMBH detections in any of the clusters, even when considering offsets from the photometric center due to Brownian motion.

\subsection{Upper Limits on Central Sources}

Because of the combined effects of Brownian motion and the uncertainties in the cluster photometric centers, we report upper limits for central point sources on the basis of the noise properties of the central region of the image as a whole, rather than a flux density limit at the location of the beam corresponding to the center.

The $3\sigma$ rms flux density in the cores of M19 and M22 is 4.5 $\mu$Jy; in M15 it is 6.3 $\mu$Jy. These limits are given in Table 1. We use these values to calculate $3\sigma$ upper limits on the masses of central IMBHs. M19 and M22 have no previously published limits on a central point source; Bash \etal~(2008) give a $3\sigma$ limit of 25.5 $\mu$Jy at 8.6 GHz for M15.

\subsection{IMBH Mass Limits}

Here we translate the flux density upper limits from $\S 3.2$ into corresponding limits on the masses of accreting IMBHs. We follow the basic method outlined in Maccarone \& Servillat (2008, see also Maccarone \etal~2005): we predict the (mass-dependent) X-ray luminosity of the IMBH, then use the radio--X-ray fundamental plane for accreting black holes in the low/hard state (Merloni \etal~2003; Falcke \etal~2004) to convert the radio constraint to a mass constraint. We briefly summarize the principal assumptions here; more details may be found in the cited references. Uncertainties in these assumptions are discussed in $\S 3.4$. 

We assume that the X-ray luminosity is given by the standard equation: $L_X=\epsilon \dot{M} c^2$ with radiative efficiency $\epsilon$ and mass accretion rate $\dot{M}$. Rather than the standard $\epsilon=0.1$, we assume that at low accretion rates the efficiency scales linearly with the accretion rate: $\epsilon \propto \dot{M}$. Here we define a low accretion rate as $\dot{M} / \dot{M}_{\rm edd} < 0.02$, which is the typical state transition luminosity for X-ray binaries (Maccarone 2003). Requiring continuity with $\epsilon=0.1$ at the transition gives $\epsilon=0.1 \,((\dot{M} / \dot{M}_{\rm edd}) / 0.02)$ at low accretion rates. We also assume that $\dot{M}$ is given by $3\%$ of Bondi rate for gas at $T=10^{4}$ K (Pellegrini 2005), noting that it is the product $\epsilon \dot{M}$ which is observable. The density of this gas is assumed to be $\rho=0.2 \,\,\rm{cm}^{-3}$ (see \S 3.4). Any accreting IMBHs in GCs will be in the low/hard state: sources accreting at higher rates would be easily detectable as luminous X-ray sources in existing surveys.

We use an updated form of the radio--X-ray BH fundamental plane from Plotkin \etal~(2012). Their ``contracted" sample is used, which includes Galactic stellar-mass BHs, Sgr A*, and some nearby low-luminosity active galactic nuclei (but excludes BL Lacs and FR I radio galaxies). This plane is given by:

\begin{align}
\textrm{log} \,L_X = (1.44 \pm 0.09) \textrm{log} \,L_R -  \\
(0.89 \pm 0.09)  \textrm{log} \,M_{\rm BH} \nonumber - (5.95 \pm 2.58)
\end{align}

\noindent
where $L_X$ is the X-ray luminosity and $L_R$ is the radio luminosity, both in units of erg s$^{-1}$. This version of the fundamental plane has a scatter of $\sigma=0.4$, corresponding to a factor of 2.5 in X-ray luminosity. The plane is defined at 5 GHz, and we assume flat radio spectra.

Given these assumptions, our $3\sigma$ flux density limits translate directly to $3\sigma$ IMBH mass limits. For M15, M19, and M22, these values are: 980, 730, and 360 $M_{\odot}$. Figure 2 illustrates these limits graphically.

We emphasize that these values---unlike some in the literature---utilize conservative assumptions about the accretion rate and efficiency. For example, mass limits based on X-ray non-detections alone have typically assumed accretion at the full Bondi rate and with a fixed radiative efficiency (e.g., Grindlay \etal~2001). This difference in approach explains why similar published constraints are often derived from X-ray and radio measurements, even though radio observations are more sensitive to the presence of IMBHs.

In M22, the 360 $M_{\odot}$ limit corresponds to $\epsilon \sim 3\times 10^{-6}$, comparable to that derived from some models of the accretion onto Sgr A* (Narayan \etal~1998; Quataert \& Gruzinov 2000), and the predicted X-ray luminosity is $9 \times 10^{29}$ erg s$^{-1}$. The associated accretion rate limit is $\sim 5\times 10^{-12} M_{\odot} \, \textrm{yr}^{-1}$, 0.1\% of the typical ongoing mass loss from a single cluster red giant (Dupree \etal~2009). 

\subsection{Mass Limit Uncertainties}

There are uncertainties in most of our assumptions about the properties of accretion onto possible IMBHs. Here we describe the characterization of these uncertainties and  Monte Carlo simulations to assess their cumulative effects.

We assume that the IMBH accretes at a fraction of the Bondi rate. The measurements of Pellegrini (2005) suggest that active galactic nuclei typically accrete at a few percent of the Bondi rate, but with a large scatter. We therefore assume a lognormal distribution for the fraction of Bondi ($f_b$) centered at $f_b = 0.03$, with $\sigma = 0.7$ (in log $f_b$),  and truncated above at $f_b = 1$. Since observations constrain the product $\epsilon \dot{M}$ rather than either quantity alone, this distribution of $f_b$ should be interpreted as including some uncertainty in $\epsilon$ as well.

The Bondi rate itself depends on the density of gas surrounding the IMBH. There are two GCs for which measurements of intracluster gas have been published: 47 Tuc and M15 (Freire \etal~2001). Both measurements are from variations in the dispersion measures of pulsars in GC cores, and are sensitive to the number density of free electrons. These estimates are $n_e = 0.07$ and $0.2\,\,\rm{cm}^{-3}$, respectively, although foreground contamination cannot be ruled out in the case of M15. These measurements of the free electron density indicate the presence of ionized gas. Theoretically, of order $1 \,\,\rm{cm}^{-3}$ of gas should be present in the core of a GC based on simple free-streaming mass loss from first-ascent red giant branch and asymptotic giant branch stars (Pfahl \& Rappaport 2001). We note that these estimates are far below the upper limits from typical integrated measurements of  \ion{H}{1} or free-free emission in GCs (e.g., van Loon \etal~2006; Knapp \etal~1996). We summarize these joint constraints with a normal distribution for $\rho$ with $\mu = 0.2$ and $\sigma = 0.1$, truncated at zero. 

The uncertainty in the fundamental plane was included as published in Plotkin \etal~(2012): $\sigma = 0.4$ in the prediction of log $L_X$. We also used the uncertainty in the cluster distances as listed in Table 1.

The distributions of $3\sigma$ IMBH mass limits produced by these simulations are lognormal, with $\sigma$= 0.39 dex. This lognormal form and scatter are set by the prior distribution for $f_b$, which is by far the dominant source of uncertainty. This can be readily seen as the inferred IMBH mass depends on the assumptions as: $M_{BH} \propto L_R^{0.38} (f_b \rho)^{-0.46}$ (Maccarone \& Servillat 2010);  the scatter in the fundamental plane itself is a minor contribution to the overall uncertainty. Because of the weak sensitivity of the IMBH mass limits to the input parameters, one would need to assume extremely low values of the accretion rate ($10^{-3}$ to $10^{-4}$ of the Bondi rate) or gas density to allow for the presence of $\sim 10^{3}-10^{4} M_{\odot}$ IMBHs in these GCs.

\section{Discussion}

There are currently no undisputed cases of IMBHs in GCs.

Until recently, the best candidate was probably the massive cluster G1 in M31. An HST dynamical estimate gave a mass of $(1.8 \pm 0.5) \times 10^4 M_{\odot}$ (Gebhardt \etal~2005), though this measurement is challenging because of the small sphere of influence of the IMBH (0.02--0.03\arcsec).  X-ray and radio emission associated with the GC were consistent with an IMBH of $\sim 5\times10^{3} M_{\odot}$ to $10 ^{4} M_{\odot}$ (Pooley \& Rappaport 2006; Kong \etal~2010; Ulvestad \etal~2007). However, new JVLA observations find no evidence of radio emission, and suggest that the earlier detection was spurious (Wrobel \etal~2012; J. Miller-Jones \etal, in preparation). The X-ray emission ($2\times 10^{36}$ erg s$^{-1}$) is consistent with that from actively accreting low-mass X-ray binaries. Thus, the sole evidence supporting an IMBH in G1 is the dynamical measurement.

A number of Galactic GCs have dynamical evidence for a central mass concentration that is often interpreted as an IMBH. These include $\omega$ Cen, for which there is an unresolved discrepancy between ground-based kinematics studies that find an IMBH of mass $\sim$ 4--5 $\times10^{4} M_{\odot}$ (Noyola \etal~2010; Noyola \etal~2008) and HST stellar proper motions that set an IMBH upper limit of $\sim$ 1--2 $\times10^{4} M_{\odot}$ (van der Marel \& Anderson 2010). Part of the problem is that there is a disagreement on the location of the center of the cluster, although this cannot account for the whole discrepancy. 

L{\"u}tzgendorf \etal~(2011) report a $2\sigma$ dynamical measurement of a $\sim 2 \times10^{4} M_{\odot}$ IMBH in the massive bulge GC NGC 6388. In the Sgr dwarf GC M54, Ibata \etal~(2009) report a central velocity dispersion peak that could be due to an IMBH of $\sim 10^{4} M_{\odot}$. These measurements can be improved by future data---especially the combination of radial velocities and proper motions---but will still be fundamentally limited by the small size of the IMBH sphere of influence.

In M15, Gerssen \etal~(2003, see also Baumgardt \etal~2003 and references therein) found marginal evidence for a $\sim 2000 M_{\odot}$ from HST/STIS spectroscopy but note that a population of ``dark remnants", such as neutron stars, is a viable explanation for their data.

A few GCs have had dynamical studies that yielded interesting upper limits on IMBH masses. Notably, in 47 Tuc, McLaughlin \etal~(2006) set a $1\sigma$ upper limit of 1000--1500 $M_{\odot}$ on the basis of HST proper motions; this cluster also has the tightest X-ray upper limit on a central IMBH (Grindlay \etal~2001); the $3\sigma$ radio limit using our assumptions is 1050 $M_{\odot}$ (Lu \& Kong 2011).

In none of these cases are the dynamical detections consistent with constraints derived from radio observations (for other papers, we translate the flux densities into IMBH mass limits using our generally more conservative formalism). In $\omega$ Cen, Lu \& Kong (2011) report a $3\sigma$ limit of 21 $\mu$Jy beam$^{-1}$ at 5.5 GHz, equivalent to an IMBH mass limit of $870 M_{\odot}$. For NGC 6388, the $3\sigma$ 5.5 GHz ATCA limit of 42.3 $\mu$Jy beam$^{-1}$ in Bozzo \etal~(2011) corresponds to an IMBH mass limit of $< 2300 M_{\odot}$. Wrobel \etal~(2011) give a $3\sigma$ VLA limit of 51 $\mu$Jy beam$^{-1}$ for M54; because of the distance of this GC, the  mass limit is larger: $\sim 4200 M_{\odot}$. And as discussed above, the new radio data in this paper constrains an M15 IMBH to $<980 M_{\odot}$ at the $3\sigma$ evel.

For $\omega$ Cen and NGC 6388 these limits are a factor of ten (or more) lower than the published dynamical masses. In the cases of M54 and M15, the discrepancy is smaller but still present. These discrepancies are not easily addressed. For example, a $2 \times 10^{4} M_{\odot}$ IMBH in $\omega$ Cen would need to be shining at $\la 10^{-9}$ of the Bondi luminosity (the luminosity for $\epsilon = 0.1$ and Bondi accretion) to be consistent with the radio limits. This is orders of magnitude lower than even famous low-luminosity active galactic nuclei such as Sgr A* (mentioned above) and M31* (Garcia \etal~2010), which have X-ray luminosities $\sim 10^{-5}$ to $10^{-7}$ of the Bondi luminosity.

Alternatively, the GCs could have dramatically lower gas densities than expected, which would be surprising given the general agreement between the Pfahl \& Rappaport (2001) model for mass loss and the existing pulsar observations. Additional gas density estimates from pulsar timing data would be extremely beneficial.

The joint constraints from previous radio data and our new deep JVLA observations provide strong evidence that either: (a) massive ($\ga 1000 M_{\odot}$) IMBHs are rare in GCs, or (b) any such IMBHs accrete at relative luminosities below that observed in even the lowest-luminosity galactic nuclei. We strongly support the acquisition of more deep radio continuum observations of Galactic GCs to improve the demographics of the cluster sample. 

\acknowledgments

NRAO is a facility of the NSF operated under cooperative agreement by AUI. LC is a Jansky Fellow of NRAO.  GRS and COH are supported by NSERC, and COH also by an Ingenuity New Faculty Award.

\begin{figure*}[!hp]
\includegraphics[width=19cm]{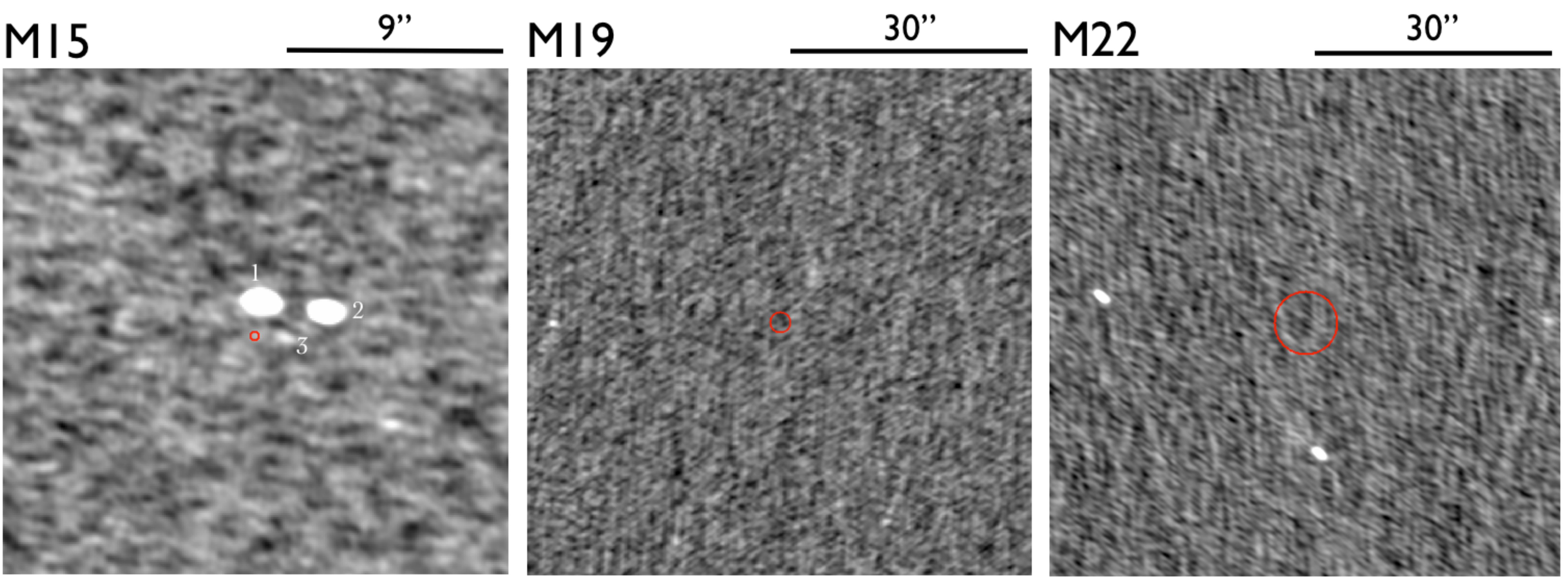}
\figcaption[ucds1]{\label{fig:thumb}
Cutout images of the cores of M15, M19, and M22. Each image has an identical stretch and contrast; the angular scales are listed. The red circles mark the 100 $M_{\odot}$ Brownian search radii (\S 3.1) around the photometric centers. Only M15 has central sources, which are all known: \#1---AC 211; \#2---M15 X-2; \#3---PSR B2127+11A.}
\end{figure*}

\begin{figure*}[!hp]
\center
\includegraphics[width=12cm]{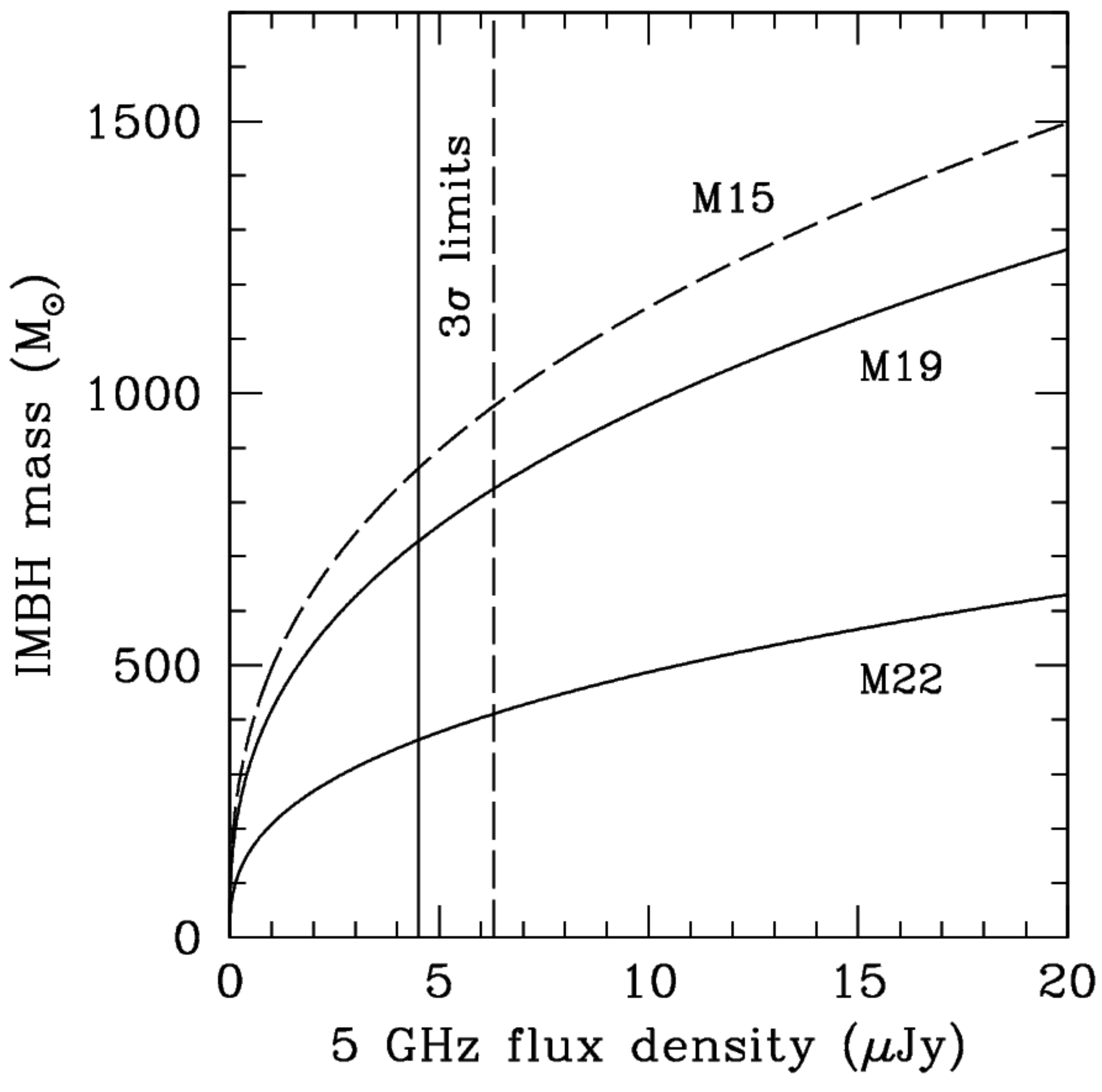}
\figcaption[ucds2]{\label{fig:thumb2}
Predicted IMBH mass as a function of 5 GHz radio flux density for M19 and M22 (solid line) and M15 (dashed line). The only difference is the cluster distance. The $3\sigma$ flux density limits of 4.5 $\mu$Jy (M19 and M22) and 6.3 $\mu$Jy (M15) are marked with the respective line styles. The intersection of the appropriate flux density limit with the model curve yields the IMBH mass limit for each cluster.}
\end{figure*}

\begin{deluxetable}{lccccccc}
\tablewidth{0pt}
\tabletypesize{\footnotesize}
\tablecaption{Globular Cluster Data \label{tab:data}}
\tablehead{ID 
& R.A. (J2000)\tablenotemark{a} 
& Decl. (J2000)\tablenotemark{a} 
& Uncertainty & Dist.\tablenotemark{b} 
& $3\sigma$ flux density\tablenotemark{c} 
& $3\sigma$ $L_{\rm R}$\tablenotemark{d}   
& $3\sigma$ IMBH mass\tablenotemark{e}  \\
& (hr:min:sec) & ($^\circ$:\arcmin:\arcsec) & (\arcsec) & (kpc) & ($\mu$Jy) & (erg/s) & ($M_{\odot}$)}
\startdata
M15 & 21:29:58.33 & +12:10:01.2  & 0.2 & $10.3\pm0.4$ & 6.3 & $4.7\times10^{27}$ & 980 \\ 
M19 & 17:02:37.80 & $-26$:16:04.7 & 1.1 & $8.2\pm0.4$ & 4.5 & $2.1\times10^{27}$ & 730 \\
M22 & 18:36:23.94 & $-23$:54:17.1 & 0.8 & $3.2\pm0.3$ & 4.5 & $3.2\times10^{26}$ & 360 \\
\enddata
\tablenotetext{a}{Source of photometric centers: M15 and M22 (Goldsbury \etal~2010); M19 (Picard \& Johnston 1995).}
\tablenotetext{b}{Source of distances: M15 (van den Bosch \etal~2006); M19 (Valenti \etal~2007); M22 (Monaco \etal~2004).}
\tablenotetext{c}{$3\sigma$ limit at average frequency of 5.9 GHz.}
\tablenotetext{d}{Equivalent $3\sigma$ radio luminosity limit.}
\tablenotetext{e}{Equivalent $3\sigma$ IMBH mass limit.}
\end{deluxetable}

\end{document}